\RequirePackage{fix-cm}
\documentclass[notitlepage]{article}
\DeclareMathAlphabet{\mathpzc}{OT1}{pzc}{m}{it}

\usepackage{mathptmx}      
\usepackage{amsmath}
\usepackage{amsfonts}
\usepackage{amssymb}
\usepackage{graphicx}
\usepackage{lineno}
\usepackage{mathrsfs}
\usepackage[mathscr]{euscript}
\usepackage{dsfont}
\usepackage{txfonts}

\def\dbar{{\mathchar'26\mkern-12mu d}}
\newtheorem{4/3 Degrees of Freedom Theorem}{4/3 Degrees of Freedom Theorem}
\newtheorem{The Natural Logarithm Theorem}{The Natural Logarithm Theorem}

\begin{document}
\bibliographystyle{plain} 
\pagestyle{myheadings}
\markright{Isotropy, entropy, and energy scaling}

\title{Isotropy, entropy, and energy scaling} 
\author{Robert Shour}\
\date{}
\begin{center}
\ {\Large\textbf\ Isotropy, entropy, and energy scaling} \\
\normalsize
\hfill \\
\ Robert Shour
\hfill \\
\hfill \\
\footnotesize{Toronto, Canada}

\normalsize
\end{center}
\hfill
\begin{abstract}
\noindent
Two principles explain emergence. First, in the Receipt's reference frame, $Deg(\mathpzc{S}) = \frac{4}{3} Deg(\mathpzc{R})$, where Supply $\mathpzc{S}$ is an isotropic radiative energy source, Receipt $\mathpzc{R}$ receives $\mathpzc{S}$'s energy, and $Deg$ is a system's degrees of freedom based on its mean path length. $\mathpzc{S}$'s $\frac{1}{3}$ more degrees of freedom  relative to $\mathpzc{R}$ enables $\mathpzc{R}$'s growth and increasing complexity. Second, $\rho(\mathpzc{R}) = Deg(\mathpzc{R}) \times \rho(\mathfrak{r})$, where $\rho(\mathpzc{R})$ represents the collective rate of $\mathpzc{R}$ and $\rho(\mathfrak{r})$ represents the rate of an individual in $\mathpzc{R}$: as $Deg(\mathpzc{R})$ increases due to the first principle, the multiplier effect of networking in $\mathpzc{R}$ increases. A universe like ours with isotropic energy distribution, in which both principles are operative, is therefore predisposed to exhibit emergence, and, for reasons shown, a ubiquitous role for the natural logarithm.  
\end{abstract}
\
\
\tableofcontents
\listoftables

\section{Introduction}
\label{intro}

This article derives two theorems. One, \textit{The $\frac{4}{3}$ Degrees of Freedom Theorem}, describes how more degrees of freedom in an energy source compared to the system receiving the energy initiates and drives emergence. The other, \textit{The Network Rate Theorem}, explains the thermodynamic benefit of networking. Together they provide a theory of emergence. Related ideas possibly explain the natural logarithm and connect quantum scaling and gravity. 

Emergence is the name assigned to a process whereby an ensemble of simple components results in a system or process with features that the components do not have. An emergent phenomenon  can not be predicted on the basis of the attributes of its fundamental components. Life, markets, language, and ecosystems are emergent systems. 

\textit{The $\frac{4}{3}$ Degrees of Freedom Theorem} is more fundamental than \textit{The Network Rate Theorem} for two reasons. First, \textit{The $\frac{4}{3}$ Degrees of Freedom Theorem} suggests how a system begins. Second, \textit{The $\frac{4}{3}$ Degrees of Freedom Theorem} can be proved with only mathematics. This article derives \textit{The Network Rate Theorem} first because it leads to \textit{The $\frac{4}{3}$ Degrees of Freedom Theorem}. \textit{The Network Rate Theorem} is revealed by an analytical approach applied to language as an emergent system. An analytical approach  attempts to account for an observed phenomenon. A synthetic approach instead would start with components and deduce an outcome based on them. A synthetic approach highlights a minimal set of relevant conditions necessary to deduce an outcome, but is hopeless as a starting point for an emergent outcome because it is impossible to know in advance what conditions are minimally relevant. ``\ldots no collective organizational phenomenon \ldots  has ever been deduced'' (Laughlin, 2005, p. 88). 

That study of a collective phenomenon like language, which emerges due to relationships among networked members of a society, could reveal laws of physics is not unexpected. The physicist David Bohm wrote: ``Einstein's basically new step [in special relativity] was in the adoption of a \textit{relational} approach to physics'' (Bohm, 1962, p. xvi). `` \ldots the physical facts concerning time and space coordinates consist only of \textit{relationships} between observed phenomena and instruments \ldots . Likewise \ldots the facts concerning perception in common experience show that this also is always concerned with relationships \ldots '' (Bohm, 1962, p. 62). The physicist Lee Smolin writes that ``networks do not exist in space---they simply are. It is their network of interconnections that define, in appropriate circumstances, the geometry of space \ldots '' (Smolin, 1997, p. 285). Robert Laughlin, who won a Nobel Prize in physics, writes ``The laws of nature that we care about \ldots emerge through collective self-organization \dots'' (Laughlin, 2005, p. xi). The philosopher David Hume wrote: ``'Tis evident, that all the sciences have a relation, greater or less, to human nature'' (Hume, 1739, p. xv). 

\textit{The Network Rate Theorem} emerges from conceptual foraging. Just as social insects follow promising paths over a physical landscape, humans follow promising paths over a conceptual landscape. Honey\-bees scout potential hive locations; their society collectively evaluates alternatives until a consensus emer\-ges (Seeley, 2010). Foraging ants follow other ants' pheromone trails, and reinforce chemical signals. Due to appraisal, choice, review and refinement, a tested idea emerges from random foraging. In connection with \textit{computerized} heuristics the mathematicians Zbigniew Michalewicz and David Fogel note: ``The essential idea of evolutionary problem solving is quite simple. A population of candidate solutions to the task at hand is evolved over successive iterations of random variation and selection. Random variation provides the mechanism for discovering new solutions. Selection determines which solutions to maintain as a basis for further exploration'' (Michalewicz, 2004, p. 161).

Four ideas are pre-eminent in this article. One: the mean path length $\sigma$ is the intrinsic scaling factor of a networked system of size $n$. Two: for $n=\sigma^\eta$ the exponent $\eta$ gives the system's intrinsic degrees of freedom and its \textit{intrinsic} entropy. Using a parameter other than $\sigma$, as in the usual definition of entropy, gives only an \textit{indirect} measure of a system's intrinsic degrees of freedom. Three: intrinsic degrees of freedom multiplies capacity. Four: in the Receipt's reference frame, an isotropic energy Supply has $\frac{1}{3}$ more intrinsic degrees of freedom than the system (Receipt) that receives the energy. 

Some observations about the mean path length, degrees of freedom, intelligence, language and mathematics follow in this Introduction. Leaving out the observations about intelligence, language and mathematics would shorten the article and avoid the distraction of possibly arguable points which do not affect the derivations. The observations are included because they provide context for reasoning leading to the derivations.

\noindent \textit{On Mean Path Lengths.} The physicist Rudolf Clausius found that oxygen molecules at the temperature of melting ice travel an average 461 meters per second  (Clausius, 1857; Brush I, p. 131). The physicist Buijs-Ballot objected that if so, ``volumes of gases in contact would necessarily speedily mix with one another''. ``How then does it happen that tobacco-smoke, in rooms, remains so long extended in immovable layers?'' (Clausius, 1858; Brush I, p. 136). In reply, Clausius introduced the concept of the mean path length (Brush I, p. 140). A gas molecule does not travel unimpeded but collides with other gas molecules. The mean path length is the average distance (some fraction of a meter) a molecule moves before its center of gravity comes into contact with the `sphere of action' of another molecule.

The psychologist Stanley Milgram (Milgram, 1967) asked, what is the length of an acquaintance chain connecting any two people selected arbitrarily from a large population. In his terminology, ``A target point is said to be of the $i^{th}$ remove if it is of the $i^{th}$ generation and no lower generation.'' Milgram asked people to mail a document towards a target in Boston. He measured the lengths of the acquaintance chains, and found a mean of 5.2 links. His experiment is the origin of the expression `six degrees of separation' presumed to separate, on average, any two people in the world. 

Clausius's mean path length can be made equivalent to Milgram's degrees of separation by equating acquaintance and collision. Instead of finding the mean distance between gas molecules in meters, find the mean number of collisions separating gas molecules. Suppose that both gases and societies collectively maximize their energy efficiency. Then the same fundamental equation that characterizes the efficient use of energy should apply to both gases and societies. In this novel way the concept of mean path length connects physical systems and networks; a general principle related to one may apply to the other.

In a seminal 1998 article, Watts and Strogatz analyzed `a small world network'. Earlier research had mostly studied entirely regular or entirely random networks. They examined randomly rewired networks of an intermediate character. They defined the `characteristic path length' as the average of the least number of steps between pairs of vertices in a graph, a definition equivalent to degrees of separation. The clustering coefficient $C$, which measures the result of their rewiring of a graph, is the network average of all $C_i$ where $C_i$  is the proportion of one step away nodes that are actually connected to each node $a_i$ in a network: $C = \frac{\sum^{n}_{i=1} C_i}{n}$. For example, suppose node $a_k$ has 5 neighbors one step away. If only 3 of them actually link to $a_k$ in one step, then the proportion  connecting in one step is $C_k = \frac{3}{5}$.  

\noindent \textit{On Degrees of freedom.} The motion of a point in a plane has two degrees of freedom. The motions of $N$ points on a plane have $2N$ degrees of freedom. One degree of freedom confers one choice on a given axis or line; a choice of  left or right does not give two degrees of freedom. 
Suppose a node moves at any given time with 3 degrees of freedom. The rate at which collisions with other nodes in the same system occur does not affect its degrees of freedom of motion at any given time. 

\noindent \textit{On the Nature of Intelligence.} IQ tests are designed and administered by psychologists. A Task Force of the American Psychological Association in 1996 characterized intelligence as the ``ability to understand complex ideas, to adapt effectively to the environment, to learn from experience, to engage in various forms of reasoning, to overcome obstacles by taking thought'' (Neisser, 1996). In this article, assume intelligence is the rate of problem solving and that solved problems can be counted. That enables mathematically modeling intelligence. Use of the model permits its appraisal. 

An IQ test indirectly measures an individual's skill, using their innate problem solving capacity, at solving the problem of learning from a society's store of solved problems and from experience, and at applying (perhaps by joining together different ideas) what that individual has learned.  Since members of a society share the same store of knowledge, average IQ  measures, partly and indirectly, the IQ of that society. Think of average individual IQ, like a collective economic indicator, as the society's average problem solving rate per capita. 

\noindent \textit{On Collective Intelligences.}  By analogy to economics, society collectively allocates its collective resources to solving problems in a conceptual area until the outcome is as beneficial as for collective resources spent on solving problems in some other conceptual area.  ``\ldots a potatoe-field should pay as well as a clover-field, and a clover-field as a turnip-field'' (the economist Jevons, 1879, p. liv). Consider problems to be the conceptual equivalents of turnips. ``The product of the `final unit' of labor is the same as that of every unit, separately considered'' (the economist, John B. Clark, 1899, p. viii): on average, solutions with the same energy cost should benefit society  to the same extent.  Just as sectors in an economy compete for financial resources, problems in a society compete for its collective problem solving resources. If an alternative use of problem solving energy gives society a better yield for its solutions, society diverts energy to that alternative use until the solution yields are about the same.

At all scales, a collective intelligence exceeds the intelligence of its component individuals. Bees, wasps, ants and termites locate, design and build nests or hives with a collective skill that exceeds the cognitive capacities of individual insects. ``Individually, no ant knows what the colony is supposed to be doing, but together they act like they have a mind'' (Strogatz 2003, p. 250). The Greek mathematician Pappus (c. 290 –- c. 350) attributed a collective mathematical insight to bees: ``Bees then, know just this fact which is of service to themselves, that the hexagon is greater than the square and the triangle and will hold more honey for the same expenditure of material used in constructing the different figures'' (Heath, Vol. 2, Ch. XIX, p. 390). Collective intelligence even occurs in bacteria (Ben-Jacob, 2010). In the field of Swarm Intelligence (SI) (Kennedy; Bonabeau), to solve difficult problems computer scientists and engineers use networked algorithms and robots to mimic the emergent collective intelligence of social insects. 

Cultures, economies and mathematics are collective intelligences. ``Although generated by the collective actions of lots of brains, cultures have storage and processing capabilities not possessed by a single human'' (Montague, 2006, p. 199), the wisdom of crowds (Surowiecki). A society's economy has ``dispersed bits of incomplete and frequently contradictory knowledge which all separate individuals possess'' (Hayek, p. 77) leading to a market solution that ``might have been arrived at by one \textit{single} mind possessing all the information'' (Hayek, p. 86). In mathematics ``the inner logic of its development reminds one much more of the work of a \textit{single} intellect, developing its thought systematically and consistently using the variety of human individualities only as a means'' (I.R. Shafarevitch, in Davis, 1995, p. 56). This also applies to  physics, literature, biology and so on.

Adoption by a society of a proposed solution to a problem depends on society's  estimate of its likelihood of success. A proposed site for a new bee hive is appraised by the old hive. The success of a new electronic device in a human society is appraised, through the operation of a market, by all potential buyers. In SI, programmers imitate this appraisal and approval effect, creating `ant pheromone trail' software  for robots. 

An ant has about one million neurons; a human brain has about 100 billion. Suppose that the same  physical laws govern the networking and problem solving output capacity of neurons in ants as in a single human brain. Then the collective behavior of 10 billion neurons in 10 thousand ants and of 100 billion neurons in a human brain should have similarities. 

Compare a society of 10,000 ants to a society of 10,000 human beings. If the same  physical laws govern, then the human society is as much more intelligent than the average  component human as is the ant society than the average component ant. A society of 100 million humans---10,000 networked societies of 10,000 individuals each---is as much more intelligent than an average society of 10,000 humans as is a society of 10,000 humans than its average component human. Consider that humanity through language, writing and culture can accumulate a store of solved problems for hundreds of human generations. The collective \textit{cumulative} intelligence of all human societies is much greater than that of an individual human. 

Bert Holldobler and Edward O. Wilson write (1990, p. 252): 
\begin{quote}
	For two reasons ants can be expected to practice economy in the evolution of their communication systems, that is, to use a small number of relatively simple signals derived from a limited number of ancestral structures and movements. First, the small brain and short life span of ant workers limit the amount of information these insects can process and store. Second, the tendency toward signal evolution through ritualization restricts the range of potential evolutionary pathways.
\end{quote}

\noindent On a different scale, the same observation applies to a human society. Consider the history of mathematics and language.

\noindent \textit{On Mathematics and language.} For this article, it is not necessary to agree on what improves mathematics and language, or how. It is only necessary to assume that language and ideas improve. Grounds for that assumption follow.

From the time of the Babylonians  5,000 years ago until now mathematics has improved in the quality and efficiency of concepts, methods and notation (Boyer, 1991; Cajori, 1928; Menninger, 1992). Like bees evaluating reports of nest sites from bee scouts, incremental improvements in problem solving by individuals are evaluated by groups of people for efficacy and efficiency. Similarly, language  ``is a continuous process of development'' (Aitchison, 1989, quoting Wilhelm von Humboldt, 1836). Historical linguistics (McMahon, 1994; Campbell, 1998) records a history of improvement in the formation of sounds and words; ``a law of economy'' (Herder, 1772, p. 164). The linguist Otto Jespersen (p. 324) observed that language ``demands a maximum of efficiency and a minimum of effort \ldots  [this] formula is simply one of modern energetics''. The linguist April McMahon writes ``\ldots sound systems tend toward economy'' (p. 30). In general, ``\ldots saving mental effort may be the most important kind of economy'' (Polya, 1962, II, p. 91). I conjecture that the average rate of progress in the efficiency of language is, for the economic reasons discussed above, the same as the rate of progress in mathematics. Both encode ideas---data.

Data compression software provides an analogy to mathematics and language. In the 1990s, as the amount of electronic data transmitted, stored and accessed increased, and the processing power of computers increased, efficient data compression became economically important. Data compression software steadily improves (Salomon, 2007).  As society's knowledge---data---increases, language increases compression of data  it encodes through naming (categorization, or unification),  contraction (can't), clipping (bike, bus, condo) (Campbell, p. 278), acro\-nyms (IBM), allusion (his, computer, next door), pattern (Salomon, p. 7), and metaphor (sunny disposition). Grammar---word order and word endings---  ``\ldots provides relief to memory'' (Diderot, The Encyclopedie). ``Declensions and conjugations are merely shortcuts\ldots'' (Herder, 1772, p. 160).  

For humans ``\ldots language first of all is classification and arrangement of the stream of sensory experience \ldots . In other words, language does in a cruder but also in a broader and more versatile way the same thing that science does'' (the linguist Benjamin Whorf, 1956, p. 55). Society tests the utility, efficacy and consistency of words that encode---compress---perceptions and ideas.  ``All knowledge is a structure of abstractions, the ultimate test of the validity of which is, however, in the process of coming into contact with the world that takes place in immediate perception'' (Bohm, 1965, p. 262).  

To encode and compress data into words requires a society to collectively solve problems (McMahon, p. 138) that include: (1) how to devise and choose sounds to be used for encoding; (2) how to assemble sounds into words; (3) what percepts and concepts should be encoded. A language embodies a set of encoding problems collectively solved by the generations of a society that use it. It is a product of collective intelligence.  Society performs the same function for language that software engineers (and their users) perform for data compression software. Hierarchies efficiently organize categories, structures, and methods for the assembly of words into larger structures such as sentences and theories. Analogy eases learning, remembering and using a language; similar endings, sounds, sentence structures, rhythms and musicality provide a `relief to memory'. The problem solving elements of language---identifying concepts and encoding them---also apply to mathe\-matics, physics, and ideas generally. People who adopt encodings with increased compression juggle more information per time unit. Speedier problem solving is of immense value to organisms with finite life spans.  

Language also improves by adding new words: encodings of new or modified concepts. Collections of concepts, such as theories, also improve at all scales.

Feedback from each use of a word is a scientific experiment. Society has tested more encoded concepts more comprehensively, in more ways, more often, for periods of time far longer, than they could be tested by any individual. Individual intelligence relies on an enormous store of highly tested and refined conceptual problem solving tools created by thousands of generations of human societies. An individual using someone else's verified solution saves energy. Through compression a language increases in depth, and through the encoding of new concepts it increases in breadth. Diderot remarked over two centuries ago that `` \ldots by merely comparing the vocabulary of a nation at different times, one would get a sense of its progress'' (Diderot).

Mathematical concepts are more efficient, compressed, defined, and have been more widely tested, than concepts encoded into words. Societies test their own language over many generations, but all societies in all cultures have tested mathematics in millions of contexts repeatedly over hundreds of generations. Mathematics can be more precisely tested than words both through logical analysis and because the accuracy of mathematics compared to physical phenomena can be \textit{measured}. ``Mathematics is a part of physics. Physics is an experimental science, a part of natural science. Mathematics is the part of physics where experiments are cheap'' (Arnold, 1997). 

Just as 10,000 ants building a nest exhibit an intelligence far greater than that of an individual ant, our store of mathematical knowledge created over the past several thousand years by tens of thousands of mathematicians, tested and appraised by tens and hundreds of millions of people in daily, scientific and commercial contexts, exhibits an intelligence beyond human comprehension. Mathematics as a disembodied network of concepts `knows' things that individuals do not. The `unreasonable' effectiveness of mathematics in the natural sciences (Wigner 1960; Hamming 1980) is like magic. An ``education in \ldots mathematics is a little like an induction into a mystical order'' (Smolin, p. 178). The difference between the intelligence of mathematics and of an individual human is far greater than the difference between the intelligence of 10,000 societies and the intelligence of an individual human.

Since mathematical ideas result from collective efforts of millions of people over hundreds of generations, common mathematical ideas such as counting numbers must reflect fundamental principles underlying the natural world. Mathematical reasoning relies on a higher, collective, intelligence. A mathematical theorem that predicts phenomena subsequently observed or that connects to other mathematical ideas is a form of experimental verification, the outcome of humanity's collective scientific evaluation of mathematical concepts.  Mathematical deduction can render express what is implicit in collective mathematical knowledge. A mathematical concept that appears to  apply to a phenomenon can be tested  by applying it in different contexts, just as society collectively does. 

\noindent \textit{On a question about intelligence.} Average IQs increase. Is this due to the improvement of language and ideas? This question impels foraging over the conceptual landscapes below. 

\section{The Network Rate Theorem} \label{SectionNetRateThm}

\subsection{Observations leading to \textit{The Network Rate Hypothesis}}

Consider words as part of society's accumulated array of problem solving tools. If a lexicon improves, so should problem solving. Researchers have observed that average IQs have increased---improved---in the U.S. in the past 60 years or so by about 3.315\% per decade; no one knows why (Flynn, 2007, p. 113, Table 1 at p. 180). Are increasing average IQs caused by words improving? Both improving at the same rate would be positive evidence. A rate characteristic of language improvement is needed. \textit{Measuring} the depth of words is difficult. Measuring the breadth (number) of words is a massive undertaking, but has already been accomplished by academic dictionaries. If word counts of a lexicon at two different times use the same criteria, and if each count is large, then the calculated rate of increase in the lexicon should be a good estimate of the rate of collective problem solving, because lexicons require a large number of problems to be collectively solved.

The English lexicon increased from 200,000 words in 1657 (Lancashire, EMEDD) to 616,500 words in 1989 (Simpson, OED), 3.39\% per decade. The University of Toronto's partly completed Dictionary of Old English (DOE) contains Old English words from the year 600 to the year 1150. Eight of  22 Old English letters, up to the letter \textit{g}, had been completed at December 2008. Extrapolating from the 12,271 words for the 8 completed letters---the dictionary counts \ae\ as a separate letter---and assuming the same average number of words per letter, gives 34,020 words in Old English for the whole Old English alphabet of 22 letters. An increase from 34,020 words in 1150 to 616,500 words in 1989 in the OED is an increase of 3.45\% per decade. Both English lexical growth rates are close to the rate at which average IQs increase. 

The error arising from using an  estimate $(\rho(Lex))_{Est}$ of the actual rate of English lexical increase $(\rho(Lex))_{Act}$ is calculable by comparing the difference between the actual size of the English lexicon in 1989, $\left[N(t_2)\right]_{Act}$ and an estimate $\left[N(t_2)\right]_{Est}$ based on an estimated English lexical growth rate $(\rho(Lex))_{Est}$ applied to an actual initial English lexicon. Set $\Delta N_2=\left[N(t_2)\right]_{Est} -\left[N(t_2)\right]_{Act}$. Then 

\begin{equation}\label{eq 17}
\\ (\rho(Lex))_{Est}-(\rho(Lex))_{Act}=\left[ \ln \left(1+\frac{ \Delta N_2}{\left[N(t_2)\right]_{Act}}\right) \right] \div \Delta t.
\end{equation}

\noindent The error in the  estimate of $(\rho(Lex))_{Act}$ becomes smaller as the time period $\Delta t$ increases.
 
Do other studies measure the rate of increase in ideas accumulated by society? The efficiency of lighting in terms of its labor cost increased from 1750 B.C.E. to 1992 by $ \frac{ 41.5}{.000119} = 348,739.5$ times according to a study by the economist William Nordhaus (1997). That is 3.41\% per decade, close to the rates for English lexical growth. Societies appraise lighting improvements and choose which ones to adopt. In light of Equation (\ref{eq 17}), a study covering 3,742 years gives a good estimate of the average rate at which ideas improve. 

Do better collective ideas increase longevity and reduce homicides? Jim Oeppen and James Vaupel (2002) found that, for example, male longevity in Norway increased from 44.5 years in 1841 to 71.39 years in 1960; that is 3.97\% per decade. Manuel Eisner (2003) estimates the London homicide rate in 1278 at about 15 per 100,000 inhabitants (p. 84) compared to the English homicide rate in 1975 of 1.2 per 100,000 inhabitants (p. 99), a rate of decrease of 3.75\% per decade. The rates based on these studies are only in the vicinity of that for increased lighting efficiency, perhaps due to the choice of data.

Morris Swa\-desh devised a method to estimate, based on the rate of their divergence, when two daughter languages had a common mother tongue. His method is called glottochronology. First he compiled a list of 100 or 200 words basic to languages (the Basic List). He then calculated the rate of change between cognates (such as moi in French and me in English) by comparing their use in historical records.  He found (Swadesh, 1971) an average rate of divergence of two daughter lexicons of about 14\% per thousand years. In 1966, he used this divergence rate to estimate that Indo-European (English's ancestral language) existed at least 7,000 years earlier (p. 84).  Gray and Atkinson (2003) dated Indo-European to 8700 years earlier, using newer methods. Updating Swadesh's calculation using Gray and Atkinson's findings, two daughter languages diverged from each other at $ \approx \frac{7000}{8700} \times 14\%=11.2\%$ per thousand years; or $5.6\%$ per thousand years \textit{each} from a notionally static mother tongue. Why half the updated Swadesh divergence rate $5.6\%$ per thousand years is so much slower than the English lexical growth rate is a new problem. Swadesh's method of estimating the divergence rate has been severely critiqued on criteria for identifying cognates and other grounds (Blust, 2000, p. 204; Campbell, 1998). 

If the updated Swadesh divergence rate estimates the common origin of two daughter languages, does a rate exist which estimates when language itself began?  To approximate the size of such a rate, suppose the 616,500 words of the 1989 Oxford English Dictionary grew from 100 words in 200,000 years. That would be $4.3\%$ per thousand years, not far off half the updated Swadesh divergence rate. Is half the updated Swadesh divergence rate a  fossil rate $\rho(\mathfrak{r})$ embedded in the much faster English lexical growth rate? This leads to: 

\noindent \textit{The Network Rate Hypothesis}: There is a function $\eta$ (small Greek eta) such that $3.39\%$ per decade in collective lexical growth = $\rho(\mathpzc{R})=\eta \times \rho(\mathfrak{r})$, where  $\rho(\mathfrak{r})$ is some kind of fossil rate. 

The significance of being able to measure the rate of improvement in \textit{collective} problem solving, via increasing average IQs, lexical growth and improvement in lighting, is that measurability converts a qualitative question---do concepts improve---into a testable hypothesis. 

\subsection{Deriving and modeling $\eta$}

To investigate how language facility increases for an individual, ask how a child acquires words. A child learns words from two parents, who each learn from  two parents, and so on. Suppose there are $\eta$ antecedent generations, and (as an idealization and simplification) each generation  \textit{independently} increases society's accumulation of words at the same rate. If parents were the only source for words, the number of word source generations would be $\log_2 (2^\eta) = \eta$. But other people can be word sources. The scaling factor (the base of the log function) is not 2, but some unknown average value $\sigma$. $\sigma$ must be determined in order to convert $\log_\sigma(n)$ into a number. What number is $\sigma$? 

Is $\sigma$ an intrinsic average number of acquaintances? Primates usually live in bands of 50 members; grooming is part of their social life (Dunbar, 1997, p. 120-122). Dunbar suggests that a person virtually `grooms' three times as many people using words as is possible grooming manually. Could $\sigma$ be 3 or 50?  Consider idealized speakers who seek to transmit information with least effort,  and idealized listeners who seek to decode information with least effort, as  distinct groups (Zipf, 1949, p. 21). Does Dunbar's optimum audience of three balance the competing goals of speakers and hearers? 

Three or fifty, these numbers cannot work. Why? Appeal to mathematical reasoning. If more persons transmit information (if $\sigma$ is greater), information received should be greater. But, on the contrary, as $\sigma$ increases, $\log_\sigma (n)$ decreases. It is impossible, if $\eta$ multiplies $\rho(\mathfrak{r})$, that the Network Rate decreases for the individual with more information sources. \textit{The Network Rate Hypothesis}, or an assumption, explicit or implicit, on which it is based, is wrong or the function $\eta$ is not logarithmic. Suppose  \textit{The Network Rate Hypothesis} is valid and that $\eta$ is a logarithmic function. Then reconsider the assumption that $\sigma$ is a fixed  number.  What \textit{parameter} would $\sigma$ have to be for $\eta$ to be logarithmic?\footnote{The balance of this article sorts out the implications of the answer to this question.} $\sigma$ must cause $\log_\sigma (n) = \eta$ to increase when $\sigma$ decreases. 

If information takes less time to reach an individual, then the rate of increase in the individual's store of information should increase. More information can be received during a lifetime. A faster average rate of information transfer implies a shorter average minimum transmission (or \textit{relationship}) distance per time unit between transmitting and receiving individuals. The mean path length corresponds exactly to such a distance. Suppose then that $\sigma$ is a network's  mean path length. For simplicity's sake, suppose that the actual number of steps between pairs of nodes equals the average number of steps, $\sigma$. Finding $\eta$ is still not complete. In an actual network not all pairs of nodes $\sigma$ steps apart are actually connected. If each node receives an average proportion  $C<1$ of the multiplicative effect of $\log_\sigma(n)$, the network needs to increase the number of nodes to $\sigma^\frac{\eta}{C}$ to have the same value of $\eta$ as a network with $C=1$. Conclude that $\eta = C \times \log_\sigma(n)$ and that in general $\rho(\mathpzc{R})  = C \log_\sigma (n) \times \rho(\mathfrak{r})$, where $\rho(\mathpzc{R})$ and $\rho(\mathfrak{r})$ are rates, and $C$ is the network's clustering coefficient. 

\noindent \textit{On the assumption that an average exists.}  To apply the formula for $\eta$ to actual networks requires that average rates proportional to the mean path length exist. Average IQs exist. Economists calculate average gross domestic product per capita. In principle criteria for counting different kinds of problems solved by people can be designed and the average number of problems solved per time period can be calculated. In principle, therefore, the average rate of problem solving per capita is calculable. If the average rate of problem solving obeys laws of economic efficiency, the average rate of lexical problem solving and the average rate of solving lighting problems in terms of labor cost, can be used as proxies for the average rate of collective problem solving. In this article, only the average features of problems are of interest. 

Obtaining a count of problems is not easy, especially counts that are reflective of society's collective problem solving (such as words in a lexicon). All that is necessary though is to assume counting is possible in principle. If problems can be counted \textit{in principle}, the average collective problem solving rate can be calculated \textit{in principle}. 

\noindent \textit{Testing $\eta$.} Before spending time and energy scouting for an explanation for the proposed $\eta$, determine if it works. If it does, then why it does will be the next problem. 

For data, researchers have measured the mean path length $\sigma$ and the clustering coefficient $C$ for some networks. Data on a line in Table \ref{Table 1} is used to calculate $\eta(n) = C \log_\sigma(n)$ for the same line. The values of $\sigma$ and $C$ for a population of actors (Watts \& Strogatz, 1998) are applied to human societies generally; this is  justified below using the \textit{Natural Logarithm Theorems} and \textit{The $\frac{4}{3}$ Degrees of Freedom Theorem}. 

\renewcommand{\arraystretch}{1.1}
\begin{table}
\begin{center}
\footnotesize
	\begin{tabular}{|c|c|c|c|c|c|c|}\hline
	\textbf{Network} & Nodes  & Number of nodes & $\sigma$ & \textit{C} & $\eta$ & Notes\\ \hline
Actors & people & 225,226 & 3.65 & 0.79 & 7.52 & {\footnotesize\ 1} \\ \hline
\textit{C. elegans} & neurons & 282& 2.65 & 0.28 & 1.62 &{\footnotesize\ 1}\\ \hline
Human Brain & neurons & $10^{11}$ & 2.49 & 0.53 & 14.71 & {\footnotesize\ 2}\\ \hline
1989 English & words & 616,500 & 2.67 & 0.437 & 5.932 & {\footnotesize\ 3, 4}\\ \hline
1657 English & words  & 200,000 & 2.67 & 0.437 & 5.431 & {\footnotesize\ 4, 5}\\ \hline
1989 population & people & 350,000,000 &  3.65 & 0.79 & 12.0 & {\footnotesize\ 6, 7}\\ \hline
1657 population & people & 5,281,347 &  3.65 & 0.79 & 9.445 & {\footnotesize\ 6, 8}\\ \hline
	\end{tabular}
	\caption{Calculations of $\eta$} \label{Table 1}
	\end{center}

\small
\noindent \textbf{Notes to Table} \ref{Table 1} \\
\footnotesize
1.\ $C$ and $\sigma$ are based on values in the article by Watts and Strogatz (1998).\\
2.\ The number of neurons: Nicholls, 2001, p. 480. $\sigma$ and $C$: Achard, 2006.\\
3.\ The number of words: OED (Simpson). \\
4.\ $\sigma$ and $C$: Ferrer, 2001 based on about 3/4 of the million words appearing in the British National Corpus. Motter (2002) found $\sigma$=3.16 and $C$=.53 based on an English thesaurus of about 30,000 words, a smaller and less representative sample.\\
5.\ The number of words: EMEDD (Lancashire).\\
6.\ $\sigma$ and $C$: based on the actors study of Watts and Strogatz (1998).\\
7.\ The number of people is an estimate of the English speaking societies in 1989, by adding  censuses: 1990  USA, 248.7 million people (Meyer, 2000); 1991 Canada 27,296,859; 1991 England 50,748,000; 1991 Australia, 16,850,540 people. These total 343,595,000 people.\\
8.\ The number of people in England: Table 7.8, following p. 207, for the year 1656, Wrigley, 1989. 
\end{table}
\normalsize

If \textit{The Network Rate Hypothesis} is valid, the average problem solving capacity $(\rho(\mathpzc{R}))_{av}$ of English speaking society, not including the effect of using language, from 1657 to 1989 is $(\eta(pop))_{av}=\frac{9.445+12.0}{2}=10.72$ times the average individual rate, $\rho(\mathfrak{r})$. Treat English society itself, without the use of language, as a single collective brain with innate problem solving capacity $(\rho(\mathfrak{r}))_{Coll}=10.72 \times \rho(\mathfrak{r})$. Multiply $(\rho(\mathfrak{r}))_{Coll}$ by the increase in capacity $(\eta(Lex))_{av}$ conferred  on $(\rho(\mathfrak{r}))_{Coll}$ by the English lexicon. For 1657 to 1989, $(\eta(Lex))_{av} = \frac{5.431+5.932}{2}=5.68$. Now find the average individual innate problem solving capacity $\rho(\mathfrak{r})$ using society's worded problem solving capacity: $\rho(\mathpzc{R}) \approx 3.41\%$ per decade $= (\eta(Lex))_{av} \times  (\rho(\mathfrak{r}))_{Coll} = (\eta(Lex))_{av} \times (\eta(pop))_{av} \times \rho(\mathfrak{r})$. 

$\rho(\mathfrak{r}) = 5.6\%$ per thousand years, exactly half the updated Swadesh divergence rate. 
\begin{quote}
	``Such an agreement between results which are obtained from entirely different principles cannot be accidental; it rather serves as a powerful confirmation of the two principles and the first subsidiary hypothesis annexed to them'' (Clausius, 1850).
\end{quote}

\noindent In this case, the subsidiary hypothesis is \textit{The Network Rate Hypothesis}\footnote{From about June 2007 to June 2009, I averaged a starting individual rate of 0 and $\rho(\mathfrak{r})$, instead of averaging the $\eta$'s of the lexicon and population and holding $\rho(\mathfrak{r})$ constant over the relevant time period. This gave $\rho(\mathfrak{r})=$ 2.8\% per thousand years, a $4 : 1$ ratio instead of the correct 5.6\% per thousand years, a $2: 1$ ratio. I persisted in studying $\eta$ because of the precision of the $4 : 1$ ratio. Some of my older preprints on arXiv have this error.}.

Using values from about 1989 in Table \ref{Table 1}, 
\begin{equation} \label{eq IndividualIQ} 
	\rho(\mathpzc{R})=\eta(pop_{1989}) \times \eta(Lex_{1989}) \times \rho(\mathfrak{r}) = 12 \times 5.932 \times \rho(\mathfrak{r}) = 71 \times \rho(\mathfrak{r}). 
\end{equation}
Equation (\ref{eq IndividualIQ}) implies that what a 1989 individual experienced as a proprietary rate of  problem solving, $71 \times \rho(\mathfrak{r})$, mostly derives from $\eta(pop) \times \eta(Lex)$.

What manner of concept is $\eta$? Why does it work? 

To simplify, assume that all binodal distances are $\sigma$ steps, all nodes have equal capacities to receive and transmit information and all transmissions have an equal amount of information and use the same amount of energy. Assume a network with $\sigma^\eta = n$ information sources. With these simplifying assumptions, the focus is on network level characteristics. Like the temperature outdoors, component level (molecular) characteristics and variations are irrelevant. One number suffices. If $\rho(\mathfrak{r})=k\sigma$, $\log_\sigma(\sigma^\eta) =\log_{k \sigma}((k\sigma)^\eta)=\log_{k \sigma}(k^\eta \sigma^\eta)$. 

Instead of two parents, four grandparents and so on supplying words, each first generation receiver  has $\sigma^2$ second generation sources, $\sigma^3$ third generation sources and so on up to $\sigma^\eta = n$  $\eta^{th}$ generation sources. \textit{Each} node receives the $\eta$ benefit of networking, which implies that all possible connections form. Then each node has $\sigma + \sigma^2 + \ldots + \sigma^\eta$ sources of information. But since the $\eta^{th}$ generation alone has $n=\sigma^\eta$ nodes as information sources, as well as information sources in generations $1$ through $\eta -1$, each node would have more information sources than there are nodes. A related issue is, suppose the network receives $n$ units of energy per time unit for each generation of information exchange involving a particular node. There is not enough energy (and therefore not enough time in a round of information transmission) for all possible combinatorial states. Can this counting problem be resolved? (This problem relates to the ergodic hypothesis, discussed below.) If not, the hypothesis fails.

Next is the commensurability problem posed by dimensional analysis (Bridgman, 1922): how can the mean path length, a measure of distance, scale $n$, a population size? A scaling subgroup for a population should be a sub-population, not a distance. Third is the $n-1$ problem: If a given node receives information from the rest of the network, consistency requires that the argument of the log function should be $n-1$ not $n$, unless the node transmits, impossibly, new information to itself. Fourth, how would such a network be wired? Fifth (a vexing problem of categorization): is a mean path length a distance or a scaling factor?

If the first four problems are irresolvable (the fifth problem will be dealt with separately later), then $\eta(n) = C \log_\sigma(n)$ must be false. Yet $\rho(\mathfrak{r})$ matches half the updated Swadesh divergence rate too closely to be coincidence. Since $\eta$ applies to transmission of information, ideas from information theory may help. 

Claude Shannon derived a formula (1948) for the information content $\eta$ of a string of 0s and 1s, where $p_i$ is the probability of the $i^{th}$ symbol,
\begin{equation}\label{eq EntropySigma}
	\eta =  \sum p_i \log_2 \left( \frac{1}{p_i} \right).
\end{equation}

\noindent Shannon used a graph to show that $\eta$ is maximum for a given number of bits when the probability of each bit occurring is the same  ($p_i = p_j, \forall i \neq j$), which is also explained (Khinchin, 1957, p 41) by Jensen's inequality. Shannon's observation is called the maximum entropy principle (Jaynes). Equation (\ref{eq EntropySigma}) has the same form as that used for entropy in thermodynamics,
\begin{equation}\label{eq EntropySigmaWK}
	K \times \sum p_i \log_x \left( \frac{1}{p_i}\right).
\end{equation}

Assume equality of all of a network's nodes, $p_i = \frac{1}{n}, \forall i$ in Equation (\ref{eq EntropySigmaWK}). Substitute $\sigma$ for $x$. Then $\sum \frac{1}{n} \log_\sigma(\frac{1}{1/n}) = \log_\sigma(n)$.  $\eta$ in  \textit{The Network Rate Hypothesis} has the same form as Equation (\ref{eq EntropySigmaWK}). $K$ in Equation (\ref{eq EntropySigmaWK}) corresponds to the clustering coefficient $C$.  ``\ldots discoveries of connections between heterogeneous mathematical objects can be compared with the discovery of the connection between electricity and magnetism \ldots'' (V.I. Arnold); connections between different mathematical models imply they share a common principle. $\eta$'s connection to entropy connects $\eta$ to thermodynamics.

Commensurability: suppose that the number of nodes $n$ and the mean path length $\sigma$ are both proportional to a common measure of \textit{energy}.  If it takes $\sigma$ energy units to travel $\sigma$ steps, then an average of $\sigma$ people are  within $\sigma$ steps of each of the network's nodes.

Counting problem: $\eta$ mathematically requires multiple scalings yet a constant argument $n$. Energy must scale in a uniformly \textit{nested} way. A cluster of nodes scales by $\sigma$, not like a pyramid, adding nodes at each next proceeding level, but internally, by uniformly subdividing into $\sigma$ subclusters. For example, 27 nodes can be internally scaled by 3 as follows:

\begin{equation}\label{Eq Internal Scaling 100.100}
\left[\left\{aaa\right\}\left\{aaa\right\}\left\{aaa\right\}\right] \ \ \left[\left\{aaa\right\}\left\{aaa\right\}\left\{aaa\right\}\right] \ \ \left[\left\{aaa\right\}\left\{aaa\right\}\left\{aaa\right\}\right].
\end{equation}

\noindent A node \textit{when networked} as in (\ref{Eq Internal Scaling 100.100}) has 3 network capacities, depending on in which size cluster, 3 nodes, 9 nodes, or 27-nodes, its capacity is exercised: $\eta_3(27) = \log_3 (27) = 3$. In a next generation, the number of clusters increases by $\sigma$ and the number of nodes per cluster decreases by $1/\sigma$. The number of nodes per generation is constant. The first generation has $\sigma$ clusters each with $\sigma^{\eta - 1}$ nodes, the second generation has  $\sigma^2$ clusters each with $\sigma^{\eta - 2}$ nodes, and so on, until the $\eta^{th}$ generation of $\sigma^\eta$ clusters with one node each. $(k+1)^{st}$ generation clusters  nest in $k^{th}$ generation clusters.

Suppose a network of $n$ equal nodes receives $n$ energy units per time unit. Per time unit, each node only has enough energy to binodally connect to one other node, not to all $\sigma$ \textit{possible} nodes. What $\eta$ multiplies is \textit{capacity}; $\eta$ measures $\mathpzc{R}$'s degrees of freedom relative to $\mathpzc{R}$'s mean path length, $\sigma$: $\eta(n)=Deg_\sigma(\mathpzc{R})$. Since each node is in each uniformly scaled nested generation, each (average) individual has the same number of degrees of freedom as the network itself. This resolves the wiring problem. It also resolves the $n-1$ problem: ``The individual agent contributes to the dynamics of the whole group (society) as well as the society contributing to the individual'' (Dautenhahn, 1999, p. 103).

The intrinsic measure of $\rho(\mathfrak{r})$ is $\sigma$; $\rho(\mathfrak{r}) \propto \sigma$. Define the \textit{intrinsic} entropy or degrees of freedom of a system $X$ with $n$ nodes as $Deg_\sigma(X)=\log_\sigma(n)$. In network theory, $\sigma$ equals the average degrees of separation. For a society with $\sigma^\eta = n$ people in it, every person has the same `relationship stride' (acquaintanceship distance), $\sigma$. Treat the start position as the first generation or stride. Then  $\eta$ strides, or degrees of freedom, spans the society. In thermodynamics, for an ideal gas $G$,  $Deg_\sigma(G) =\log_\sigma(n)$ where $\sigma$ is the \textit{intrinsic} mean path length for colliding molecules, measured in collision steps. In information theory, $\sigma =2$ when information is represented in bits. 

I propose to categorize the hypothesis as 
\textit{The Network Rate Theorem} (\textit{NRT}): For an isotropic system $\mathpzc{R}$ with $n=\sigma^\eta$ nodes and mean path length $\sigma$
\begin{equation}\label{Eq NRHypToTheorem}
\begin{split}
	\rho(\mathpzc{R}) &  = \eta(\mathpzc{R}) \times \sigma \\
	&= Deg_\sigma (\mathpzc{R}) \times \sigma
	\end{split}
\end{equation}
\noindent and in general for a clustering coefficient $0<C \leq 1$
\begin{equation}\label{Eq NRHypToTheorem200}
	\rho(\mathpzc{R})  = C \log_\sigma(n) \times \rho(\mathfrak{r}).
\end{equation}

\noindent When $C=1$ uniformly scaled nested clusters model $\mathpzc{R}$.

\subsubsection{\textit{The NRT}: Observations, implications, and speculations}

\textit{On the special role of the mean path length.} Is there a way, without using Shannon's graph or Jensen's inequality, to show how the mean path length optimizes $\eta$? Consider a system of water containers. Level 1 has $\sigma$ water containers, each supported underneath at level $2$ by $\sigma$ water containers.  Water is supplied at the same rate to each of the first level water containers. When a first level water container is full, water spills into its supporting level $2$ water containers. If one level $2$ container is smaller than the rest which are equal in size, it spills water while the other containers are still filling. If one level $2$ container is bigger than the rest which are equal in size, the rest spill water while the bigger container is still filling. Analogize water to energy. A networked system utilizing energy supplied at a fixed rate will increase its rate of output if it uses more (and wastes less) of the energy supplied per time unit. Nested, uniformly scaled distribution of energy from a Supply $\mathpzc{S}$ induces a nested, uniformly scaled structure in a networked system $\mathpzc{R}$ receiving the energy, as otherwise energy supplied per time unit by $\mathpzc{S}$ is not fully utilized by $\mathpzc{R}$. 

Suppose a central energy source radiates energy. Efficient flow must be uniform in every direction. A wave front circular from the source maximizes entropy.

What mechanism allows a network to find its average scaling factor? Unite the concept of an ideal network  with the concept of an ideal heat engine. The idealized network discussed above consists of all possible pairs of nodes, all $\sigma$ steps apart and equal in capacity. In Sadi Carnot's (1824) ideal heat engine the cylinder contains a working substance such as air between a fixed plate and a movable frictionless piston. A furnace transmits heat to the otherwise perfectly insulated cylinder, causing the working substance to expand. The furnace ceases contact with the cylinder and is put in touch with a heat sink which removes heat from the working substance causing it to contract. Then the heat sink is removed, and the cycle repeats.  The piston cycles up and down moving an attached articulating arm. Carnot proved that no heat engine can be more efficient than an ideal heat engine. No energy is lost other than to moving the articulating arm.

Consider the piston's initial position and the unique turning point in the heat cycle to be two nodes: a heat engine's heat cycle is intrinsically binodal. Treat the furnace as one node and the heat sink as the other. Remove the articulating arm. Place a furnace and a heat sink at each node, so that energy can equally well move from one node to the other. A binodal symmetric ideal heat engine can perfectly transfer energy from one node to the other. 

Suppose that the amount of energy required to transmit information is proportional to the amount of information transmitted. By analogy, construct a binodal symmetric ideal information engine which transmits information from one node to the other. All nodes have identical transmission and reception capacities with no energy or information loss. Form an ideal network consisting of symmetric ideal information engines. No information exchange network can be more efficient. Each generation of isotropic information exchange is equally and perfectly efficient. If the physical environment changes, a network whose nodes all have equal capacities in each generation of information exchange will be the quickest to cycle through the generations of information exchange required to reach an optimal fitness landscape. This (I conjecture) models how networks binodally communicate change to their constituent components.  \textit{Optimal local binodal exchange leads to global optimality}; social insects are an example of ``a decentralized multiagent system whose control is achieved through locally sensed information'' (Kube \& Bonabeau, 2000, p. 91), as are languages (speakers and hearers), markets (buyers and sellers), and genes (two strands of DNA).

\noindent \textit{Comparing entropy and intrinsic entropy.} In 1848, William Thomson (Lord Kelvin) used Sadi Carnot's analysis of an ideal heat engine and the contraction of gases when cooled to find absolute zero (Kelvin, 1848). The volume of an ideal gas contracts in proportion to absolute temperature. Clausius sought and found an invariant property of the ideal heat engine cycle. He called it entropy (Clausius, 1865, p. 400; in English translation, 1867,  p. 365). In Clausius's derivation of entropy (1879, p. 79), he compares the volumes of the working substance at different stages of the heat cycle and finds (p. 83) that

\begin{equation}\label{Eq Entropy100}
	\frac{{\dbar Q_1}}{T_1} - \hspace{.1cm}\frac{ {\dbar Q_2} }{T_2} =0 ,
\end{equation}
where \hspace{.1cm}$ \dbar$ is an inexact differential, \hspace{.1cm}$\dbar Q_1$ is the heat added to the heat engine from the furnace at the absolute temperature $T_1$, and \hspace{.12cm}$ \dbar Q_2$ is the heat removed from the heat engine by the heat sink at the absolute temperature $T_2$.  \hspace{.1cm}$\frac{\dbar Q}{T}$ is the change in entropy. 

Boltzmann (1872) remarked that a system can achieve equilibrium on a macroscopic scale. For example, air has a measurable temperature. At a microscopic scale, on the other hand, there is constant molecular motion. He inferred that the \textit{average} exchange of energy of gas molecular collisions must also be steady. ``The determination of average values is the task of probability theory'' (p. 90, English translation). Boltzmann's \textit{H} Theory ($-H=$ entropy) used probability and a log function. Building on Boltzmann's work, the physicist Max Planck  derived the formula for entropy $\eta=K \sum  p_i \ln \left( \frac{1}{p_i} \right)$ (Planck, 1914). 

Clausius's definition of entropy is baffling. $\frac{{\dbar Q}}{T}=dS$ is the change in entropy, but what does it represent? The ratio definition, based on an ideal heat cycle, relies on experiment. A degree Kelvin equals a degree centigrade based on the freezing and boiling points of water. 0 degrees Kelvin was determined by experiment. The important practical advantage of Clausius's ratio definition is its use of temperature, which can be easily measured. 

In \textit{The Network Rate Theorem} $\rho(\mathpzc{R}) = \eta \times \sigma \Leftrightarrow \eta = \frac{\rho(\mathpzc{R})}{\sigma}$, where $\sigma$ is the system's mean path length. Changing a system's entropy changes its degrees of freedom. For example, $d \eta = \log_\sigma (\sigma^m)- \log_\sigma(\sigma^n) = m-n$. Assume a system's output rate $\rho(\mathpzc{R})$ equals its energy input rate $\rho(\mathpzc{E})$. Then, $\eta = \frac{\rho(\mathpzc{E})}{\sigma}$. In Clausius's definition of entropy, $\frac{\dbar Q}{T} = \frac{d \mathpzc{E}}{\epsilon}$, $\mathpzc{E}$ being a total amount of energy and energy ${\epsilon}$ a scaling factor. The numerator on the left side $\frac{\dbar Q}{T}$ is a change in heat, which is equivalent to a change in energy, and the denominator is the absolute temperature, which is proportional to an amount of energy $\epsilon$. Clausius's definition of entropy is equivalent to  $\eta = \frac{\rho(\mathpzc{E})}{\sigma}$, except that it uses a scaling factor $T$ \textit{proportional to} $\sigma$ in the denominator. Clausius's ratio definition so \textit{indirectly} measures a system's intrinsic degrees of freedom that it altogether obscures its connection to degrees of freedom. 

Replace $\eta$ in \textit{The Network Rate Theorem} by $\log_\sigma(n)$, and 
\begin{equation}\label{Eq NRTandBoltzmann}
	\rho(\mathpzc{E})=\rho(\mathpzc{R}) = \log_\sigma(n) \times \sigma.
\end{equation}
\noindent A mathematical union of a ratio definition of \textit{intrinsic} entropy with the statistical definition of \textit{intrinsic} entropy gives \textit{The Network Rate Theorem}. 

If two gases at different temperatures mix, they will reach an equilibrium state with a common temperature. A calculus proof of this uses differential equations. More simply, when two gases mix, repeated binodal collisions lead to a new mean path length $\sigma$ for the combined system, and hence a common average temperature ($\propto \sigma$). 

Mechanics studies how two particles interact. It is not possible to consider every collision, for example, of 6.02 times $10^{23}$ oxygen gas molecules (about 32 grams worth). Boltzmann had the idea of dividing a space up into cells, and calculated the expected statistical distribution of energies among the different cells. Trillions of molecules have a small set of different speeds or energies, a statistical mechanics. Using the mean path length to scale a system reduces the number of parameters from a small set to one, a conceptually compressed statistical mechanics. Like categorizing a country's wealth by its GDP per capita. 

\noindent \textit{On degrees of freedom and system capacity.} The exponent of a system's mean path length $\sigma$ in $n= \sigma^\eta$ measures its intrinsic degrees of freedom. The collective rate of a system $\rho(\mathpzc{R}) = \log_\sigma(n) \times \rho(\mathfrak{r})$, where  $\rho(\mathfrak{r})$ is the rate of an average individual. Suppose that $\rho(\mathfrak{r})$ is a constant, as is the case for average innate human problem solving capacity over the past few thousand years. While average individual innate capacity is unchanging, average individual capacity increases if the individual's innate capacity has more degrees of freedom to which it can be applied. That occurs when an individual adds to their store of solved problems---knowledge. 
In cellular phone networks, researchers observe that increasing the degrees of freedom in multiple input multiple output (MIM0) antenna systems leads to a `gain' in capacity (Jafar, 2008; Borade, 2003; Molisch, p. 521). More antennas, more degrees of freedom. 

\noindent \textit{On glottochronology: reconciling the divergence rate with the  English lexical growth rate.} $5.6\%$ per thousand years,  half the updated Swadesh divergence rate, equals the innate individual average problem solving rate. A population $M$ with a common mother tongue divides into daughter populations $D1$ and $D2$ isolated from each other, each initially with the same lexicon as $M$. Assume that $D1$, $D2$ and $M$ all have the same size populations and same size lexicons: at $t_0$, $Lex_M =Lex_{D1} = Lex_{D2}$, and $pop_{D1} = pop_{D2}= pop_{M}$, so $\eta(pop) \times \eta(Lex) = \eta$ for $D1$, $D2$ and $M$. To find one half the average divergence rate assume $Lex_{M}(t_1)=Lex_{M}(t_0)$, so $(\rho(\mathfrak{r}))_M=0$. Then compare the rate of change for \textit{each} daughter language to the rate for a static mother language.

\begin{equation}\label{eq Lex-10.20}
\begin{split}
	 \frac{Lex_{D}(t_1)}{Lex_{M}(t_1)}
&= \frac{(1+ (\rho(\mathfrak{r}))_{D}) \times \eta \times Lex_{D}(t_0)}{(1+  (\rho(\mathfrak{r}))_{M}) \times \eta \times Lex_{M}(t_0)}  \\
&= \frac{(1+  (\rho(\mathfrak{r}))_{D})}{1+ 0}  \\
& =1+  \rho(\mathfrak{r}).
\end{split}
\end{equation}

\noindent The $\eta$'s and the lexical sizes in numerator and denominator of Equation (\ref{eq Lex-10.20}) cancel. If the daughter tongues undergo changes independent of each other, then each $Lex_D$ grows at the same average rate $\rho(\mathfrak{r})$ away from $Lex_{M}$.  The average rate of lexical divergence of the two daughter languages equals $2 \rho(\mathfrak{r})$. Swadesh's updated divergence rate (remarkably) indirectly measures twice the average innate individual human problem solving rate. 

\noindent \textit{On mitochondrial Eve.} Using maternal mitochondrial DNA, Rebecca Cann, Mark Stone\-king and Allan Wilson dated a single woman ancestor to 200,000 years ago (Cann, 1987). Nested scaling implies that dates a first generation, not an individual (Gould, 2002).

\noindent \textit{On nested scaling and the natural logarithm.} Clusters of size $\sigma$ scale by $\sigma$, so $\frac{d\sigma}{dt} = \sigma $. This implies $\sigma = e$. In Table \ref{Table 1}, the human brain, neurons in C. elegans, and English words all have a path length close to  $e \approx 2.71828$, leading to a conjecture: 

\begin{The Natural Logarithm Theorem} The natural logarithm is a consequence of uniformly nested energy scaling. 
\end{The Natural Logarithm Theorem}

\noindent The number of generations is proportional to time. The ubiquitous role of the natural logarithm in dating processes evidences the uniform nested scaling of isotropic systems. 

\noindent \textit{On economics.} An objection to applying statistical mechanics to economics is `individuals are not gas molecules' (Sinha, 2011, p. 147). The mean path length is a bridge. Let $(\rho(EcGr))_{av}$ be a country's average rate of economic growth and $(LP)$ its labor participation rate. Let the economic productive capacity of the average working individual equal their \textit{economic} problem solving capacity $(\rho(\mathfrak{r}_{Ec}))_{av} \approx$ 3.41\% per decade. To mathematically model economic growth  (a problem in economics (Helpman)) using  \textit{The Network Rate Theorem}, 

\begin{equation} \label{EqEcGrowth100}
	(\rho(EcGr))_{av} = (\eta(pop))_{av}  \times(LP)_{av} \times (\rho(\mathfrak{r}_{Ec}))_{av}. 
\end{equation}

\noindent The average individual in a society has the same number of degrees of freedom in their rate of economic problem solving $(\rho(\mathfrak{r}_{Ec}))_{av}$ as the entire society, which at any given time is $\eta(pop) \times \eta(K)$, where $K$ represents society's  store of knowledge. It follows  that 

\begin{equation} \label{EqEcGrowth200}
	(\rho(EcGr))_{av} = ((\eta(pop))_{av})^2 \times (\eta(K))_{av} \times (LP)_{av} \times \rho(\mathfrak{r}),
\end{equation}

\noindent where $\rho(\mathfrak{r})$ is the average innate problem solving rate. If education increases $(\rho(\mathfrak{r}_{Ec}))_{av}$ in Equation (\ref{EqEcGrowth100}), economic growth should increase.

Data can test Equation (\ref{EqEcGrowth100}). $(LP)$ is about $66\%$ for the U.S. (Mosisa). Now estimate the U.S. economic growth rate from 1880 to 1980. In 1880 the U.S. had 50,155,783 people (1880 census, Table Ia) and in 1980, 226,545,805 (1980 census, Table 72). $\eta(pop_{1880})=10.8186$ and  $\eta(pop_{1980})=11.7386$, so $(\eta(pop))_{av}=11.4851$. Then
\begin{equation}\label{EqEcGrowth300}
\begin{split}
 \rho(\mathpzc{EcGr})&= 11.4851 \times 0.66 \times 3.41\% \ per \ decade \\
&=2.53\% \ per \ year.
\end{split}
\end{equation}

U.S. productivity per hour from 1880 to 1980 increased about 2.3\% per year (Romer, 1990) close to the calculated rate in Equation (\ref{EqEcGrowth300}).

Morality and laws might arise as an emergent set of rules for protecting the factors on the right side of Equation (\ref{EqEcGrowth100}). Utility theory applied to economic maximization is challenged by problems like choosing whether to hurt a person to save several people. The dichotomy of utility theory is built into Equation (\ref{EqEcGrowth100}). 

\noindent \textit{On cosmology.} Suppose the universe is 13.7 billion years old (about $4.32\times 10^{17}$ seconds) with constant scaling factor $s$ proportional to time. Suppose its entropy is $10^{123}$ (Frampton, 2008). For \textit{The Network Rate Hypothesis}, let $\rho(\mathpzc{S})$ be the age of the universe. Then
\begin{equation}\label{Eq ValueOfQuantum}
	\rho(\mathpzc{S}) = 4.32 \times 10^{17} \ seconds \ =\eta \times s= 10^{123} \times s,
\end{equation}
\noindent which implies that $s$ has a finite quantum size proportional to about $10^{-105}$ of a second, much smaller than one Planck time, about $10^{-43}$ seconds. Perhaps intriguing.  

\noindent \textit{On possible connections to quantum mechanics.} $\eta$ repetitions of $\sigma$ is wave-like. Discrete clusters are  particle-like. Nestedness of cluster generations resembles superposition in quantum mechanics. Clusters are countable like quanta. In quantum mechanics, $\mathpzc{E} = h\nu$, where $\mathpzc{E}$ is energy, $h$ is Planck's constant, and $\nu$ (small Greek nu) is frequency. Is this an analog of \textit{The Network Rate Theorem} ($h$ being the analog of the scaling factor)? 

Hugh Everett  (1957) discussed a `many worlds' interpretation of quantum mechanics. Nested degrees of freedom can replace `many worlds'.

\noindent \textit{Robots and algorithms.} Nested scaling or degrees of freedom of robots and algorithms should increase the efficiency of such systems. 

\noindent \textit{On epidemiology.} Transmission of disease is analogous to transmission of information. If $\rho( \mathpzc{r})$ can be determined, then it may be possible to calculate $\rho(\mathpzc{R})$ for a population. 

\noindent \textit{NRT testing.} Scaling occurs in allometry. Does \textit{The Network Rate Theorem} apply there?

\section{The $\frac{4}{3}$ Degrees of Freedom Theorem}

Allometry is the study of scaling relationships in organisms. 

In the allometry of metabolism $Y = a M^b$; $Y$ is the organism's metabolism, $a$ is a constant, and $M$ is the organism's body mass. In \textit{The Network Rate Theorem}, the exponent of the scaling factor varies with size, for metabolism the exponent $b$ is fixed. \textit{The Network Rate Theorem} must be adapted to apply it to allometry. 

First, some background. In 1879, Karl Meeh supposed that $b = \frac{2}{3}$: an organism's surface area dispersing body heat grows by a power of 2 while its mass supplying heat grows by a power of 3 (Whitfield). But $\frac{2}{3}$ is wrong. Not all energy goes to heat; energy is also used for movement, problem solving, growth and reproduction. Kleiber's data (1932) supports $b = \frac{3}{4}$. West, Brown and Enquist (WBE 1997) compared scaling factors, an idea adapted in the derivation below. Treating the circulatory system as a transport system for materials, they found $b = \frac{3}{4}$.  Kozlowski \& Konarzewski (2004 and 2005) identified errors in WBE's mathematics; $b = \frac{3}{4}$ has not been mathematically proven.

The erroneous but usefully simpler $\frac{2}{3}$ scaling hypothesis reveals a way to adapt \textit{The Network Rate Theorem} to metabolic scaling.  Let $\rho (\mathfrak{s})$ be the  \textit{average} heat supply rate of an organism cell, and $\rho (\mathfrak{r})$ be the \textit{average} heat dispersing  rate of a unit area on its surface. Assume heat generated is proportional to a small organism's volume $V_1$ (a Supply $\mathpzc{S}$) and all generated heat is uniformly dispersed  through its surface area $\theta_1$ (a Receipt $\mathpzc{R}$). Let $V_1 \propto (\ell_1)^3$, where $\ell_1$ is a length. Scale $\mathpzc{S}$'s length by $s$. For a larger organism $V_2 \propto (\ell_2)^3= (s \ell_1)^3 = s^3 \ell_1$. Surface area $\theta_{2} \propto (\ell_{2})^2=\sigma^2\ell_1$ where  $\ell_1$ scales by $\sigma$. In general, for $\mathpzc{S}$ 

\begin{equation}\label{EqSource2-3-100}
	V_{k+1} \propto s^{3k} \ell_1 \rho (\mathfrak{s})
\end{equation}

\noindent and for $\mathpzc{R}$,

\begin{equation}\label{EqSource2-3-200}
	\theta_{k+1} \propto \sigma^{2k} \ell_1 \rho (\mathfrak{r}).
\end{equation}

\noindent By \textit{The Network Rate Theorem}, the ratio of the capacities of Supply $\mathpzc{S}$ to Receipt $\mathpzc{R}$ is

\begin{equation}\label{EqSource2-3-300}
	\frac{\log_s(s^{3k}) }{\log_\sigma(\sigma^{2k})} = \frac{3}{2}
\end{equation}. 

\noindent and of $\mathpzc{R}$ to $\mathpzc{S}$ is $\frac{2}{3}$, the ratio of 2 dimensions to 3. In $\frac{3}{4}$ metabolic scaling, Supply $\mathpzc{S}$ is the circulatory system, Receipt $\mathpzc{R}$ is the organism, and the ratio of  their dimensions is 1. How is a $\frac{4}{3}$ ratio possible? 

A connection to $\frac{4}{3}$ occurs in an intermediate step in the proof of Stefan's Law (Allen \& Maxwell, p. 742--743; Longair, 2003, p. 256--258) concerning isotropic energy radiation. Boltzmann derived Stefan's empirically determined law (Boltzmann, 1884). Planck has the intermediate step as (1914, Ch. II, p. 62)

\begin{equation}\label{Eq PlanckOnStefansLaw}
\left( \frac{\partial S}{\partial V} \right)_T = \frac{4u}{3T},
\end{equation}

\noindent where $S$ is entropy, $T$ absolute temperature, $V$ volume, and $u=\frac{U}{V}$ is energy density. $U$ is the total energy of the system. 

The left side of Equation (\ref{Eq PlanckOnStefansLaw}) is the change of entropy per change in volume at absolute temperature $T$. Since an ideal gas volume changes in proportion to $T$, the left side measures how entropy changes relative to a scaling factor, $\partial V$, proportional to $T$.  $\frac{u}{T}$ on the right side measures the number of scalings in $u$ (energy density received) based on $T$. Hence, implicitly Equation (\ref{Eq PlanckOnStefansLaw}) says that for uniformly radiating energy the number of scalings on the left side is $\frac{4}{3}$ those on the right side, a  ratio that connects to metabolic scaling. 

If $\frac{4}{3}$ scaling applies to radiating energy then $\frac{4}{3}$ scaling should apply at all scales. Isotropic radiation explains the sphericity of the space fillers used in WBE 1997: $Deg(sphere) = 3$. Isotropy also is a large scale feature of the universe (Fixsen, 1996). Connect isotropy at all scales to energy distribution in organisms. Treat $\mathpzc{S}$ as uniformly scaled and nested. For the circulatory system, the aorta is the first generation and the capillaries are the $\eta^{th}$. Identify the average radius for a part of a cone of radiation with the radius of a tube to obtain:
\hfill \\

\noindent \textbf{The $\frac{4}{3}$  Degrees of Freedom Theorem}: In $\mathpzc{R}$'s reference frame, $Deg(\mathpzc{S}) = \frac{4}{3} Deg(\mathpzc{R})$, where $\mathpzc{S}$ is  an isotropic supply of energy and $\mathpzc{R}$ receives $\mathpzc{S}$'s energy.

\noindent \textbf{Proof:} The uniformly nested scaled model must be extended to account for an initial energy source. An initial energy source is external to a system. A system's degrees of freedom are within it. Hence, a $0^{th}$ generation energy supply is required. Designate a point source $0$ as the $0^{th}$ generation of an energy supply $\mathpzc{S}$. 

Uniformly scaled nesting or degrees of freedom corresponds to $\mathpzc{S}$ isotropically radiating energy  at all scales at $\rho(\mathfrak{s})=s$ energy units per time unit into a Receipt $\mathpzc{R}$. Consider $0$ and all points in a cone of radiated energy originating from it as comprising a Supply. Let $\ell$ represent a radial distance traveled by radiation at the rate $s$ energy units per time unit or scaling. Let $V_k$, $\forall k \geq 1$, be the portion of the cone ($V_k$ contains sub-Supplies) with average radial length $\ell_k$. (The radial length is  averaged since the ends of $V_k$ are curved surfaces.) 

Let $s$ scale $V_k$, such that $V_{k+1}= sV_k$. Since energy density $D_{k+1} = \frac{1}{s} D_k$, $\rho(\mathpzc{E}_{k+1}) =\rho(V_{k+1}) \rho(D_{k+1}) = s\rho (V_k) \frac{1}{s} \rho(D_k) =\rho(V_k) \rho(D_k) = \rho(\mathpzc{E}_k$); the rate of radiation is constant.

To be able to compare the number of degrees of freedom in $\mathpzc{S}$ relative to $\mathpzc{R}$, let $\gamma$ `scale' the average radial length $\ell_k$ of $V_k$:  $\gamma \equiv \ell_{k+1}/\ell_k = 1$. From $0$ to the far end of $V_{k+1}$, the radiation front has  $\eta(\ell_{k+1}) = \log_\gamma(\gamma^k) = k$ scalings; the radiation front is $(k+1) \times \ell$ from $0$.  In $\mathpzc{S}$, $Deg_s(s^k V_1) = Deg_\gamma(\gamma^k\ell_1)=Deg_s(s^k\ell_1)$ and if $Deg(V)=1$, then $Deg(\ell)=1$. In $\mathpzc{S}$, $\gamma=s$.

Let $r_k$ be the average radius of $V_k= \pi (r_k)^2 \ell_k$. Cones have a uniform slope. Let $\beta \equiv r_{k+1}/r_k$ represent the scaling of the average radii for $V_k$.  

Scaling factors $s$, $\beta$ and $\gamma$ are instrumental variables for comparing the degrees of freedom in Supply $\mathpzc{S}$ relative to the degrees of freedom in its Receipt $\mathpzc{R}$. 

In  $\mathpzc{S}$, since $\ell_i =\ell, \ \forall i>0$,
\begin{equation}\label{Eq4-3Scaling200}
	\frac{V_{k+2}}{V_{k+1}} = s = \frac{s^{k+1}V_1}{s^k V_1} = \frac{\pi (\beta ^{k +1 }r_1)^2 ( \ell_{k+2})}{\pi (\beta ^{k }r_1)^2 (\ell_{k+1})}= \beta^2,
\end{equation}
so $\beta = s^{\frac{1}{2}}$. If $Deg_s(V_k)=1$, then $Deg_\beta(r_k) =\frac{1}{2}$.

Since radiation is isotropic, for every $V_k$ transmitting energy at the rate $\rho(\mathpzc{E}_k)$ let $\theta_k$ be a corresponding spherical Receipt receiving energy at the same rate and scaling by $\sigma$ with radius $\xi_k = \frac{1}{2} \ell_k$ scaling by $ \alpha \equiv \xi_{k+1}/\xi_k $.  $\sigma$ and $\alpha$ are instrumental variables for determining, in $\mathpzc{R}$, the degrees of freedom of the sphere $\theta_k$ in $\mathpzc{R}$ relative to $\theta_k$'s radius $\xi_k$. If $Deg_\sigma(\theta_k)=3$, then $Deg_\alpha(\xi_k)=1$ and 
\begin{equation}\label{Eq4-3Scaling100}
	\frac{\theta_{k+2}}{\theta_{k+1}} = \sigma = \frac{\sigma^{k+1}\theta_1}{\sigma^k \theta_1} = \frac{\frac{4}{3}\pi (\alpha ^{k +1}\xi_1)^3}{\frac{4}{3}\pi (\alpha ^{k }\xi_1)^3}=\alpha^3,
\end{equation} 
so $\alpha = \sigma^\frac{1}{3}$. If $Deg_\sigma(\theta_k)=1$, then for $\xi_k= \frac{1}{2}\ell_k$ in $\mathpzc{R}$  $Deg_\alpha(\xi_k)= Deg_\gamma(\ell_k) =\frac{1}{3}$. But in $\mathpzc{S}$ $Deg_\gamma(\ell_k)$ would be $1$.

Compare the relative number of degrees of freedom of $V_k$ and $\theta_k$. Use the relationship between the radius $\xi_k$ of $\theta_k$  and the average radial length $\ell_k$ of $V_k$: $\xi_k = \frac{1}{2} \ell_k$.  In $\mathpzc{S}$, $Deg_s(\ell)=1$. Since $\xi_k$ in $\mathpzc{R}$ has $\alpha = \sigma^\frac{1}{3}$, in $\mathpzc{R}$ $Deg_s(\ell)=\frac{1}{3}$. In the third line of Equation (\ref{Eq4-3Derivation}), $\gamma$ cannot have both $1$ and $\frac{1}{3}$ degrees of freedom in terms of $s$. Calculating in the third line the relative number of degrees of freedom of $V_k$ scaling by $s^1$ in $\mathpzc{S}$ \textit{compared to} $\theta_k$ scaling by $\sigma^1$ in $\mathpzc{R}$ requires specifying in which reference frame, $\mathpzc{S}$ or $\mathpzc{R}$, the calculation is taking place. 

In $\mathpzc{S}$'s reference frame in the first two lines, and in $\mathpzc{R}$'s reference frame with $\gamma = s^{\frac{1}{3}}$ in the third and fourth lines:
\begin{equation}\label{Eq4-3Derivation}
\begin{split}
	 V_{k+1} &= s^k V_1 \; \; \; \; \; \; \; \;\; \; \; \; \; \; \; \; \; \;\;\; \; \; \; \; \; \; \;\; \;\; \;\; \;(in \ \mathpzc{S})\\
	 &= \pi (\beta)^{2k} (r_1)^2 (\gamma^k) \ell_1 \; \; \; \; \; \; \; \;\; \; (in \ \mathpzc{S})\\
	 &= \pi (s^{\frac{1}{2}})^{2k} (s^{\frac{1}{3}})^k (r_1)^2 \ell_1 \; \; \; \; \;  (in \ \mathpzc{R})\\
	 &=\pi s^{\frac{4}{3}k}(r_1)^2 \ell_1. \; \; \; \; \; \; \; \;\; \; \; \; \; \; \; \; \; \;\; (in \ \mathpzc{R})
\end{split}
\end{equation} 
When $s$ scales $V_k$ in $\mathpzc{S}$, $\sigma^{\frac{4}{3}}$ scales $\theta$ in $\mathpzc{R}$, so in $\mathpzc{R}$ $Deg_s(\mathpzc{S}) = \frac{4}{3} Deg_\sigma(\mathpzc{R})$. The extra $\frac{1}{3}$ degree of freedom of $\mathpzc{S}$ in $\mathpzc{R}$'s reference frame is due to the effect of radial motion in $\mathpzc{S}$. QED.\\

\subsubsection{\textit{The $\frac{4}{3}$ DFT}: Observations, implications, speculations}

Comments below about energy, quantum mechanics and gravity are speculations.   

\noindent \textit{On metabolic scaling.} Assume that for organisms $k$ and $k+1$, their masses $M_k<M_{k+1}$, and that every organism $\mathpzc{R}$  isotropically receives energy from a circulatory system with energy supply capacity  $\mathpzc{S}$. 

Assume that $\mathpzc{R}$'s circulatory system volume $V$  is proportional to $\mathpzc{R}$ 's volume $\theta$. By \textit{The $\frac{4}{3}$ DFT}, in $\mathpzc{R}$, $\mathpzc{S}_{k+1} \propto V_{k+1}=s^{\frac{4}{3}}V_k \propto s^{\frac{4}{3}}\theta_k$, since $V \propto \theta$.  Assume that $\mathpzc{R}$'s average number of cells  $N \propto M$, its mass, and that $M \propto \theta \propto \rho(\theta) = Y$, its metabolism. Then, for a Receipt $\theta_{k+1} = s \theta_k \propto s M_k$. 

The exponent of the $s$ factor of the Supply $s^\frac{4}{3}V_k$ must be $1$ to match the degrees of freedom of $M_k$. A $\frac{3}{4}$ power of the Supply's volume $V_{k+1}$,

\begin{equation} \label{EqEnergySupply100.100}
	(V_{k+1})^{\frac{3}{4}} = (s^\frac{4}{3}V_k)^{\frac{3}{4}}= s(V_k)^{\frac{3}{4}} \propto s( \theta_k)^{\frac{3}{4}} \propto s(M_k)^{\frac{3}{4}}= (s^\frac{1}{3}(sM_k))^{\frac{3}{4}}=(s^\frac{1}{3})^{\frac{3}{4}} (M_{k+1})^{\frac{3}{4}}
\end{equation}
  
\noindent supplies energy at the (Receipt) rate $Y_{k+1}$ to the Receipt $M_{k+1}$. Hence, when $V \propto \theta$, the energy supplied by $V$ is proportional to $M^{\frac{3}{4}}$ so $Y \propto M^{\frac{3}{4}}$. 

Different mathematical reasoning gives the same result and implies that $\frac{3}{4}$ metabolic scaling occurs at the cellular level. Let $\rho(\mathfrak{r})$ be an organism's average cellular metabolic rate. Modify \textit{The NRT} by adding factors $x$, $y$ and $Deg_{m}( \rho (\mathfrak{r}))$, $m$ a scaling factor:

\begin{equation} \label{EqMetabolicSc100.100}
	\rho (\mathpzc{R}) =x Deg_{\sigma} (\mathpzc{R}) \times y Deg_{m} ( \rho (\mathfrak{r})) \times \rho (\mathfrak{r}).
\end{equation}
 
\noindent \textit{The $\frac{4}{3}$ DFT} implies $x=\frac{4}{3}$ in Equation (\ref{EqMetabolicSc100.100}). Assume that $N$ does not vary (which can be shown to imply $V \propto M$). That is the idealized case for a mature organism. Then in Equation (\ref{EqMetabolicSc100.100}) $\mathpzc{R} \propto \theta$, $\rho (\mathpzc{R})$, and $Deg_{\sigma} (\mathpzc{R})$ are constants; $ \rho (\mathfrak{r})$ as an average is constant.  Since $\rho(\mathpzc{R}) = Deg_\sigma(\mathpzc{R}) \times \rho(\mathfrak{r})$ by \textit{The NRT}, $y$ in Equation (\ref{EqMetabolicSc100.100}) must be $\frac{3}{4}$. Scaling up of an organism's energy supply \textit{capacity} is offset by scaling down of its average cellular \textit{rate} of energy use. Thus the \textit{metabolic capacity}, which is the product $\frac{4}{3}Deg_\mathpzc{R} (\mathpzc{R}) \times \frac{3}{4}Deg_{\rho (\mathfrak{r})} ( \rho (\mathfrak{r}))$, is invariant: $\frac{4}{3} \times \frac{3}{4}=1$. This observation uses the fact that for $\rho(\mathfrak{r})=k\sigma$, $\log_\sigma(\sigma^\eta) =\log_{k \sigma}((k\sigma)^\eta)$.

The metabolism of an organism's $N$ cells is  $Y= \rho(\mathfrak{r}) \times N$, $N$ times average cellular metabolism. Apply Equation (\ref{Eq4-3Derivation}) and the value of $y$ in Equation (\ref{EqMetabolicSc100.100}). Then
\begin{equation}\label{EqMetabolicSc100.500}
\begin{split}
	\rho(\mathpzc{R}_{k+1}) = Y_{k+1} & =\sigma^{\frac{4}{3}} N_k \times m^{\frac{3}{4}} \rho(\mathfrak{r}_{k}).
	\end{split}
\end{equation}

If in Equation (\ref{EqMetabolicSc100.100}) $y$ instead equals $1$, then $\theta_{k+1}=\sigma^{\frac{4}{3}}\theta_k$. Space expands.

\noindent \textit{$\frac{4}{3 }$ DFT and economics.} For the same reason as in metabolic scaling, in economic enterprise there are economies of scale. On the other hand, increasing the efficiency of individuals frees up energy that can increase $\eta$, boosting economic growth.

\noindent \textit{Turnstile Analogy.} A stadium has four seating levels, each with rows of $n$ seats. It has three exit levels each with $n$ turnstiles. Each stadium level empties one row per time unit. Each turnstile level only allows a maximum of one row to exit per time unit. If the stadium is full, the emptying rate of the stadium levels is $\frac{4}{3}$ the exiting capacity of the exit turnstiles. Two possible remedies are: (1) increase the number of turnstiles (the Receipt) by $\frac{1}{3}$; (2) decrease the rate of exiting persons by $\frac{3}{4}$. The second solution applies to metabolic scaling. The first solution is consistent with the expansion of the universe. That is, if $\rho(\mathfrak{r})$ in Equation (\ref{EqMetabolicSc100.100}) does not scale down, then $\theta$ must scale up. 

\noindent \textit{A theory of emergence.} Together, \textit{The $\frac{4}{3}$  Degrees of Freedom Theorem} and \textit{The Network Rate Theorem} explain emergence. In $\mathpzc{R}$'s reference frame, $\mathpzc{S}$ with $\frac{1}{3}$ more degrees of freedom than $\mathpzc{R}$ initiates $Deg(\mathpzc{R})$ and causes $Deg(\mathpzc{R})$ to increase. That increases the multiplicative effect of $\eta$ in \textit{The Network Rate Theorem}. Structures (stars, and organisms) and processes (ecosystems, languages, markets, and mathematics) emerge at all scales.

Michael Stumpf and Mason Porter recently (2012) suggested that allometric scaling for metabolism has, of all putative scale-free power laws, the most evidentiary support. (The ratio, $\pi$, of the circumference to the diameter of a circle is an example of a scale-free \textit{relationship} in Euclidean geometry.) The metabolism power law is a manifestation of \textit{The $\frac{4}{3}$ Degrees of Freedom Theorem}, which may be the universe's most fundamental scale-free power law. If the universe is finite, then there are smallest and largest scales for \textit{The $\frac{4}{3}$ Degrees of Freedom Theorem}. 

$\mathpzc{S}$ scaling creates space in $\mathpzc{R}$. $\mathpzc{R}$ having been created, $\mathpzc{S}$ increases its degrees of freedom to fill $\mathpzc{R}$. Perhaps a push-pull mechanism makes time one directional. 

The fractal dimension of iso\-tro\-pic energy distribution (the Supply) is $(\frac{4}{3})^{rd}$ that of the Receipt. Fractality of a Supply $\mathpzc{S}$ induces fractality in its Receipt $\mathpzc{R}$ at all scales.

\noindent \textit{Another natural logarithm theorem.} Isotropy suggests the following. 

	\begin{The Natural Logarithm Theorem} For a finite isotropic network $\mathpzc{R}$, the base of the logarithmic function  describing $\mathpzc{R}$'s intrinsic degrees of freedom is the natural logarithm.\\
\noindent \textbf{Proof:} The contribution of networking to the multiplication of capacity per transmitting node of $\mathpzc{R}$'s $n=\sigma^\eta$ nodes as a proportion of $\eta$ is
\begin{equation}\label{Eq SM-Ec 300.20}
\begin{split}
	 \frac{d \eta}{dn}&=\frac{ d \left[\log_{\sigma}\sigma^{\eta}\right]} {d(\sigma^{\eta})}\\
&=\frac{1}{\ln\left(\sigma\right){\sigma^{\eta}}}.
\end{split}
\end{equation}
The per node reception of the increase in capacity $\eta$ due to networking as a proportion of $\eta$ is $\frac{1}{n}= \frac{1}{\sigma^\eta}$. For an isotropic network, the contribution to the increase in capacity per transmitting node, as described in Equation (\ref{Eq SM-Ec 300.20}), equals the increase in capacity per receiving node, so 
\begin{equation}\label{Eq SM-Ec 400.20}
\frac{1}{\sigma^{\eta}} =\frac{1}{\ln\left(\sigma\right){\sigma^{\eta}}}\Rightarrow \ln(\sigma)=1\Rightarrow \sigma = e.
\end{equation}
\end{The Natural Logarithm Theorem}

An information network where every node has an equal capacity to transmit and receive is isotropic.  Dunbar's optimal audience of three is slightly more than $e \approx 2.71828$ hearers. 

\noindent \textit{On Clausius's $\frac{3}{4}$ Mean Path Length Theorem.} Clausius in his paper introducing the concept of mean path length (Clausius, 1858, p. 140 of translation in Brush) notes:

\begin{quotation}
The mean lengths of path for the two cases (1) where the remaining molecules move with the same velocity as the one watched, and (2) where they are at rest, bear the proportion of $\frac{3}{4}$ to $1$. It would not be difficult to prove the correctness of this relation; it is, however, unnecessary for us to devote our time to it.
\end{quotation}

\noindent One can prove Clausius's theorem using \textit{The NRT} and \textit{The $\frac{4}{3}$ DFT}. 

\noindent \textbf{Theorem}: The mean path length of an isotropic Supply $\mathpzc{S}$ is $\frac{3}{4}$ of the Receipt's ($\mathpzc{R}$'s). 

\noindent \textbf{Proof:} Per time unit, every $k^{th}$ generation Supply $\mathpzc{S}_k$ isotropically supplies $\mathpzc{E}_k$ energy to a corresponding Receipt $\mathpzc{R}_k$. If no energy is lost in transmission, then $\rho(\mathpzc{E})=\rho(\mathpzc{S})=\rho(\mathpzc{R})$. Let $\mathpzc{S}$'s mean path length $s$ scale $\mathpzc{S}$ and $\mathpzc{R}$'s mean path length $\sigma$ scale $\mathpzc{R}$. Then, 
\begin{equation}
	\begin{split}
	\rho(\mathpzc{E}) &= \rho(\mathpzc{S}) \\
	& = Deg_s(\mathpzc{S}) \times s \; \; \; \; \; \; \; \;\; \; \left(Network \ Rate \ Theorem \right)\\
	& = \left( \frac{4}{3} \right) Deg_\sigma(\mathpzc{R}) \times s \; \; \; \; \; \; \; \;     \left( \frac{4}{3}  \ Degrees \  of \ Freedom \  Theorem \right)	\\
	& = Deg_\sigma(\mathpzc{R}) \times \left( \frac{4}{3} \right) s \\
	& = Deg_\sigma(\mathpzc{R}) \times \sigma \; \; \; \; \; \; \; ( \rho(\mathpzc{S}) = \rho(\mathpzc{R}))\\
	& = \rho(\mathpzc{R}) \; \; \; \; \; \; \; \;\; \; \; \; \;\; \; \; \; \; \;\; \; \left(Network \ Rate \ Theorem \right).
	\end{split}
\end{equation}
It follows that $\frac{4}{3}s =  \sigma$ and so $s = \frac{3}{4} \sigma$. QED.
\\

The mean path length of a social network---a Receipt---receiving isotropically  transmitted information  should be $\frac{4}{3}$ of $e$ ($=2.71828$), about $3.624$. This is close to the 3.65 found by Watts and Strogatz for actors (Table (\ref{Table 1})). Using current values for the mean path length for the 1657 population and lexicon in Table (\ref{Table 1}) is justified since those values are close to what theory suggests they should be.

\noindent \textit{A speculation about energy.} Use dimensional analysis. A cluster in a generation is analogous to a mass (from a particle point of view). Let one cluster width also be one unit of distance (from a wave point of view). The number of clusters in a generation is a distance per unit of $\eta$. Since $\eta$ is proportional to time, the mass of clusters in a generation per time is $(mass)(length)/(time)$ . $\mathpzc{S}$ radiates one generation per unit of $\eta$ or time,  $(length)/(time)$. The dimension of a  radiating mass of clusters per generation is

\begin{equation} \label{EqDimAnalEnergy100}
	 (mass)(length)/(time)\times (length)/(time) = (mass)(length)^2/(time)^2. 
\end{equation}
 
\noindent Energy has the same dimensions as the right side of Equation (\ref{EqDimAnalEnergy100}). That  implies that energy is due to the excess  $\frac{1}{3}Deg_s(\mathpzc{S})$ in $\mathpzc{R}$'s reference frame. A $0^{th}$ generation (the sun) supplies energy to planets circling it, a black hole supplies energy to stars circling it, and a singularity supplies energy for the universe that emerges from it. 

\noindent \textit{On Huygens principle.} The physicist Christiaan Huygens in 1690 had the idea that ``every point on a propagating wavefront serves as the source of spherical secondary wavefronts'' and ``the secondary wavelets have the same frequency and speed'' (Hecht, 2002, p. 104), consistent with uniformly scaled nested degrees of freedom.

\noindent \textit{On dimensions.} In $\mathpzc{S}$, $Deg(\mathpzc{S})=3$ compared to a radial length's $Deg =1$. In $\mathpzc{R}$, $Deg(\mathpzc{R})=3$ compared to a radial  length's $Deg =1$.  In $\mathpzc{R}$'s reference frame, $Deg(\mathpzc{S})=\frac{4}{3}Deg( \mathpzc{R})$. In $\mathpzc{R}$'s reference frame,  assign $\mathpzc{S}$ a $4^{th}$ dimension. Time as a $4^{th}$ dimension appears in the special theory of relativity. Time in $\mathpzc{R}$ may be due to radiative motion in $\mathpzc{S}$.

\noindent \textit{On the ergodic hypothesis.}  To derive the $H$ \textit{Theorem}, Boltzmann assumed that every combinatorial state (phase) was equally likely and would occur eventually---the ergodic hypothesis. The physicist Shang-Keng Ma comments (p. 442):

\begin{quote}
	This argument is wrong, because this infinite long time must be much longer than $O(e^N)$, while the usual observation time is $O(1)$.
\end{quote}

\noindent ($O(1)$ means the order of magnitude of a counting number. In principle there is not enough time for the ergodic hypothesis to be true.) Theory now \textit{postulates} equal probability for all points in phase space. Neither the hypothesis nor a postulate is necessary. All that is necessary is the capacity to binodally connect or collide. Then the network is scaled by its mean path length $\sigma$ in time proportional to $\log_e(O(e^N))=O(1)$. The network's degrees of freedom are equally available in an isotropic system. 

\noindent \textit{The equipartition theorem.} In Statistical Mechanics the equipartition theorem (which also relies on the equal probability of all points in phase space) provides:

\begin{quote}
In the mean each degree of freedom of the system at a temperature $T$ has the thermal energy $\frac{1}{2}kT$ (Greiner, 1995, p. 198),
\end{quote}

\noindent where $k$ is Boltzmann's constant and $T$ is absolute temperature. The ratio of 3 degrees of freedom in $\mathpzc{S}$ to 2 degrees of freedom in motion through a plane in $\mathpzc{R}$ is $\frac{3}{2}$, or $\frac{1}{2}kT$ per $\mathpzc{S}$'s degrees of freedom, a conjectural explanation. 

\noindent \textit{On a similarity to Schr\"odinger's Equation}.  A succeeding generation in a uniformly scaled $\mathpzc{S}$ has moved one $s$-scaling that is orthogonal to the preceding generation. Let $\psi$ represent a function that counts intrinsic degrees of freedom.  Bearing in mind that $\eta$ appears to be proportional to time $t$ (in $\mathpzc{R}$'s reference frame),
\begin{equation}\label{Eq Schr-Cluster}
 \psi(k+1)  = i\times s \times \frac{d \psi(k)}{d\eta}.
\end{equation}
\noindent Apply Equation (\ref{Eq Schr-Cluster}) to $k$ scalings of $s$ along all radii from a 0. Then the $k+1^{st}$ generation makes a right angle to the $k^{th}$ generation, because of the factor $i$ in Equation (\ref{Eq Schr-Cluster}), and forms a ring $k+1$ radial scalings from 0. Equation (\ref{Eq Schr-Cluster}) resembles Schr\"odinger's wave equation. 

\noindent \textit{On special relativity and quantum scaling.} In Hermann Bondi's `$k$ calculus' (1962), time for one of two inertial travelers is scaled by $k$ relative to the other. In Bondi's one dimensional paradigm (Bondi, p. 102), Brian and Alfred are on unaccelerated paths in space and pass by each other at point $0$. Alfred sends a radar pulse at time $t$ which reaches Brian at time  $kt$. Brian's response to Alfred's signal reaches Alfred at time $k(kt)=k^2t$. Bondi finds a formula for $k$ (p. 103) and  derives the Lorentz transformations that characterize the principle of relativity for inertial reference frames.

Generalize for a finite universe: for all pairs of inertial paths find the 0 common to all. Each pair of paths has a $k$ factor. The smallest is  a quantum scaling factor.

Instead of connecting points on a pair of intersecting inertial paths with straight lines as Bondi does, connect them with part of the arc of a circle. Consider a set of radial lines intersecting at $0$ with concentric circles superimposed. $t_{i+1}=kt_{i}=k^{i}t_1$. This resembles the model used in \textit{The $\frac{4}{3}$ Degrees of Freedom Theorem}. Suppose that every local $0$ acts as a gravity point relative to the set of inertial lines emanating from it. A quantum scaling factor helps describe the geometry of $\mathpzc{S}$, and is connected to gravity.

An event perceived by two inertial observers with a common $0$ as different in $\mathpzc{R}$ occurs in the same generation in $\mathpzc{S}$. The perceived relativity of time and space may occur due to the reference frame adopted. We perceive space as a single reference frame.  \textit{The $\frac{4}{3}$  Degrees of Freedom Theorem} suggests space has two. They are difficult to distinguish because they each have 3 dimensions within their own domain. 

\noindent \textit{On the Arrow.} Aristotle describes Zeno's \textit{Arrow} paradox in his \textit{Physics}: ``\ldots if everything when it occupies an equal space is at rest, and if that which is in locomotion is always occupying such a space at any moment, the flying arrow is therefore motionless''. An inertial object is not moving relative to its cluster in $\mathpzc{S}$. Inertial motion may be the perception in $\mathpzc{R}$ of $\mathpzc{S}$'s background moving relative to the inertial object\footnote{Scotty in the 2009 Star Trek movie: Imagine that! It never occurred to me to think of SPACE as the thing that was moving!}. 

\noindent \textit{The two slit experiment.} Light waves traveling through two parallel slits in a thin plate appear to interfere; if light consisted of particles, the impression left by the particles should add. Compare this to stereoscopic vision. Two eyes provide stereoscopic depth perception. Similarly,  two slits may enable  stereoscopic depth perception in $\mathpzc{R}$ of $\mathpzc{S}$'s uniform nested scaling; the observer sees the moving background as waves with amplitudes of nested heights. Perhaps the two slit experiment is a reference frame problem.

\noindent \textit{On Entanglement.} Consider a scaled generation of clusters: 
\begin{equation}\label{Eq scaling10.100}
 0 \overbrace{\longrightarrow s_1\longrightarrow s_2\longrightarrow s_3 \dots \longrightarrow s_\eta}^{s} .
\end{equation}
In (\ref{Eq scaling10.100}) a node in $s_1$ connects to a node in $s_\eta$ in $s$ steps \textit{and} in $\eta\times s$ generations or steps, counting the first generation as a step. In (\ref{Eq scaling10.100}) the number of scalings out of $\eta$ scalings is like a distance and is also a proportion of $s$. In $\mathpzc{S}$'s reference frame, from the perspective of the average scaling factor $s$:
\begin{equation}\label{eq Multiplier.10.3000}
\begin{split}
 s \supset \left\|a_{1}, b_{1}\right\|_{1} \supset \left\|a_{2}, b_{2}\right\|_{2} \supset \ldots \supset \left\|a_{\eta}, b_{\eta}\right\|_{\eta}, 
\end{split}
\end{equation}
\noindent where $A \supset B$ means $A$ contains $B$, and $\left\|a_{i}, b_{i}\right\|_{i}=\left\|a_{i}, \ldots , b_{i}\right\|_{i}$ is a representative cluster in the $i^{th}$generation. 

In $\mathpzc{R}$'s reference frame, distance is proportional to the number of scalings by $s$:
\begin{equation}\label{eq Multiplier.10.2999}
\begin{split}
 \left\|a_{1}, b_{1}\right\|_{1}+\left\|a_{2}, b_{2}\right\|_{2} + \ldots +\left\|a_{\eta}, b_{\eta}\right\|_{\eta} = \eta \times s. 
\end{split}
\end{equation}

In $\mathpzc{R}$'s reference frame, the distance $\eta \times s$ spans the system, but in $\mathpzc{S}$'s reference frame, $s$ as the average least distance between pairs of nodes spans the system. Is $s>0$ a distance or a scaling factor? Suppose the answer is, both. 

	\begin{The Natural Logarithm Theorem}

If $s$ spanning the network in one generation ($d\eta$) in $\mathpzc{S}$ is equivalent to $\eta$ segments each of $s$ units spanning the network in $\mathpzc{R}$, then $s=e$. 
\textbf{Proof:} The equivalence reduces to a differential equation per unit of $\eta$:
\begin{equation}\label{EqEntropy 100.100}
	\frac{ds}{d \eta}= \eta \times s.
\end{equation}
\noindent Consider $s$ in Equation (\ref{EqEntropy 100.100}) as if it is a function. The solution for Equation (\ref{EqEntropy 100.100}) is 
\begin{equation}\label{EqSolution100.300}
	s=e^{\eta}.
\end{equation}
\noindent If $\eta=1$, then $s=e$. Equivalence of (\ref{eq Multiplier.10.3000}) and (\ref{eq Multiplier.10.2999}) leads to the natural logarithm.
	\end{The Natural Logarithm Theorem}

In other words, the natural logarithm is evidence of space's dual reference frames. Wave particle duality 
may also be. 

Aspect, Dalibard and Roger (1982) performed an experiment that precluded communication between separated particles, and found entanglement: the result is consistent both with quantum mechanics and non-locality (Einstein, 1935; Bell, 1964). This may result from the equivalence of (\ref{eq Multiplier.10.3000}) and (\ref{eq Multiplier.10.2999}).

\section{Discussion}

The section on \textit{The Network Rate Theory} found an innate rate of lexical change that enables dating the beginning of language. If the same physical principles apply to diverse phenomena, then generalizing the same method should lead to a general theory of emergence. As conjectured above. 

The universe consists of many kinds of structures and processes, complex at all scales. One mechanism for creating complexity is many rules applied to simple components. Another mechanism is to have a simple initiating process, such as isotropic radiation or scaling, with an enormous number of degrees of freedom. If the two theorems described above are valid, they are consistent with the secondly described mechanism. 

That statistical mechanics first dealt with gas molecules is perhaps an accident of history. Statistical mechanics might also have developed by asking how much intelligence a collective intelligence contributes to its component intelligences.




\end{document}